# Why Cognitive Science is Needed for a Viable Theoretical Framework for Cultural Evolution

Liane Gabora (liane.gabora@ubc.ca)
Department of Psychology, University of British Columbia, Kelowna BC, V1V 1V7, CANADA

## Abstract

Although Darwinian models are rampant in the social sciences, social scientists do not face the problem that motivated Darwin's theory of natural selection: the problem of explaining how lineages evolve despite that any traits they acquire are regularly discarded at the end of the lifetime of the individuals that acquired them. While the rationale for framing culture as an evolutionary process is correct, it does not follow that culture is a Darwinian or selectionist process, or that population genetics provides viable starting points for modeling cultural change. This paper lays out step-by-step arguments as to why a selectionist approach to cultural evolution is inappropriate, focusing on the lack of randomness, and lack of a self-assembly code. It summarizes an alternative evolutionary approach to culture: self-other reorganization via context-driven actualization of potential.

## Introduction

Though several of the deepest evolutionary thinkers of the 20th Century cautioned against the over-zealous application of Darwinian theory (Claidière, Scott-Phillips, & Sperber, 2014; Fracchia & Lewontin, 1999; Mayr, 1996; Tëmkin & Eldredge, 2007), Darwinian models are rampant in the social sciences. The frameworks of population genetics has been applied to cultural evolution (Boyd & Richerson, 1988; Brewer et al., 2017; Cavalli-Sforza & Feldman, 1981; Creanza, Kolodny, & Feldman, 2017; Henrich et al., 2016), as well as to archaeology (O'Brien & Lyman, 2000), economics (Essletzbichler, 2011; Hodgson, 2002; Nelson & Winter, 2002), neuroscience (Edelman, 2014), the evolution of languages (Fitch, 2005; Pagel, 2017), and the unfolding of a creative idea in the mind of an individual (Campbell, 1960; Kronfeldner, 2014; Simonton, 1999; for counter-arguments see Gabora, 2007, 2011a). This paper focuses exclusively on the question of whether cultural evolution is Darwinian. This is a different project from that of examining how natural selection has shaped the propensity for culture, language, artifacts, and so forth; it models cultural change itself as a second Darwinian process.

The rationale is that since cultural forms, like biological forms, evolve, i.e., exhibit cumulative, adaptive, open-ended change, culture constitutes a second evolutionary process. This is undoubtedly true. However, cultural Darwinism goes further than the claim that culture evolves; it assumes that the formal framework of population genetics, with appropriate tinkering to accommodate culture-specific phenomena, provides a viable foundation for modeling this second evolutionary process.

Many have laid out the similarities and differences between biological and cultural evolution (Godfrey-Smith, 2012; Jablonka & Lamb, 2014; Mesoudi, 2007; Wagner & Rosen, 2014). The issue addressed here is not how similar they are, but the extent to which the algorithmic structure of cultural evolution merits importation of a Darwinian framework. This paper lays out two arguments against this project, breaking them down step by step so as to facilitate the identification and settling of any points of disagreement. The first, the weaker argument, pertains to the issue of randomness. The second pertains to the existence of a self-assembly code. We will see that due to limited interaction with cognitive science, cultural evolution research has paid little attention to structure of the human minds that evolve culture, and the processes by which elements of culture take form. This has led to the misapplication of evolutionary concepts to culture, resulting in lack of appreciation of its essentially non-Darwinian character. The paper concludes with a brief discussion of an alternative, non-Darwinian evolutionary framework for culture.

## Definitions

It is true that any definition of a term is fine so long as everyone agrees how it is being used. However, part of why it has been difficult to nail down the extent to which cultural forms evolve in the same sense as biological forms is that, in drawing parallels between biological and cultural evolution, existing terms have been stretched beyond their conventional meanings. When they are used in ways that do not capture the deep structure or essence of their original meaning, or when a biological referent is misleadingly retained in a cultural context, misunderstanding can result. The matter is tricky, for although cultural evolution constitutes a separate evolutionary process with its own evolving structures and processes, it is inextricably interwoven with biological evolution[1]. To maintain clarity, key terms used in this paper are defined below:

**Acquired trait:** a trait obtained during the lifetime of its bearer (e.g., a scar, a tattoo, or a memory of a song) and transmitted horizontally (i.e., laterally).[2]

**Culture:** extrasomatic adaptations—including behavior and artifacts—that are socially rather than sexually transmitted.

---

[1] For example, maternal diet during lactation can influence food preferences in offspring (Bilkó et al., 1994).

[2] These are acquired traits with respect to biological evolution. It will be argued that with respect to cultural evolution *all* traits are acquired.

**Darwinian process:** an evolutionary process that occurs through natural or artificial selection.[3]

**Darwinian threshold:** transition from non-Darwinian to Darwinian evolutionary process (Woese, 2002; Vetsigian, Woese, & Goldenfeld, 2006).

**Evolutionary process:** a process that exhibits cumulative, adaptive, open-ended change.

**Gene:** a component of a self-assembly code, i.e., a unit of heredity.[4]

**Generation:** a single transition period from the internalized to the externalized form of a trait.

**Horizontal transmission:** The spread of a trait within a generation.

**Inherited trait:** a trait (e.g., blood type, or the capacity to suntan) that is transmitted vertically (e.g., from parent to offspring) by way of a self-assembly code (e.g., DNA).

**Modern Synthesis:** merging of Darwin's theory of natural selection and Mendelian genetics in the 1940s.

**Organism:** the living expression of a particular self-assembly code, sometimes referred to as an 'individual'.

**Phylogenetic network:** model of the relationships amongst variants that is pictured as *reticulate,* or 'network-like'.

**Phylogenetic tree:** model of the relationships amongst different species that is pictured as *branching,* or 'tree-like'.

**Population genetics:** branch of biology central to which is a mathematical theory of how organisms evolve through natural selection due to changes in gene frequencies. It was originally developed by Fisher (1930), Wright (1931), and Haldane (1932) and subsequently expanded (Hartl & Clark, 2006).

**Selection:** differential replication of randomly generated heritable variation in a population over generations such that some traits become more prevalent than others. Selection may be *natural* (due to non-human elements of the environment) or *artificial* (due to human efforts such as selective breeding), and it can occur at multiple levels, e.g., genes, individuals, or groups (Lewontin, 1970).

**Selectionist process:** like the term 'Darwinian process' this refers to an evolutionary process that occurs through natural or artificial selection. (It will be used from this point on because it avoids potential confusion caused by the fact that Darwin considered other possibilities.)

**Self-assembly code:** a set of self-replication instructions.

**Self-other Reorganization (SOR):** a theory of how both culture, and early life, evolve through communally exchanging, self-organizing networks that generate new components through their interactions. Based on post-Modern Synthesis theory and findings in biology.

**Vertical transmission:** The inheritance of a trait from one generation to the next by way of a self-assembly code.

It is important to point out that we are using the term 'selection' in its technical, scientific sense. The word 'selection' also has an 'everyday' sense in which it is synonymous with 'choosing' or 'picking out'. One could say that selection—in the everyday sense of the term—occurs in a competitive marketplace through the winnowing out superior products. However, the discussion here concerns whether selection *in the scientific sense of the term* is applicable to cultural evolution.

Note that, with respect to biological evolution, a new generation (one transmission event) generally (though not in horizontal gene transfer) begins with the birth of an organism. It is not impossible for the same trait to be transmitted horizontally in one generation and vertically in another. However, with respect to cultural evolution, *a new generation begins with the expression of an idea* (again, one transmission event). Thus, over the course of a single discussion, an idea (a cultural trait) can undergo many generations. It can be said that cultural evolution proceeds more quickly than human biological evolution[5], since the lengthy period we associate with biological generations, from birth through development to puberty and reproductive maturity to parenthood, is in general significantly longer than the stretch of time between when an individual acquires a cultural trait (e.g., an idea) and then expresses (their own version of, or their own take on) that cultural trait.

Note also that vertical and horizontal transmission must be defined *with respect to the relevant evolutionary process*. A common error is to refer to the transmission of cultural information from parent to offspring as vertical transmission (e.g., Cavalli-Sforza & Feldman, 1981). The parent-child relationship is *with respect to* biological evolution; they are parent and child with respect to their status as biologically evolving organisms, but with respect to their status as participants in cultural evolution, there is no basis for this parent-child distinction. Indeed, while childbirth entails one mother and one father, there is no limit to the number of 'parental influences' on the 'birth' of an idea.

A related error is to say that in cultural evolution there is a third form of transmission, *oblique transmission*, in which traits are transmitted from non-kin members of the parental generation (e.g., Cavalli-Sforza & Feldman, 1981), for as far as *cultural* evolution is concerned it is irrelevant whether the information comes from *biological* kin or non-kin.

In a similar vein, although *dual inheritance* theorists speak of culture as a second form of inheritance (Henrich & McElreath, 2007; Richerson & Boyd, 1978; Whiten, 2017; Müller, 2017), the distinguishing feature of an inherited trait is that it is transmitted vertically—e.g., from parent to

---

[3] Although evolution by selection is the process Darwin's name became most synonymous with, it is interesting to note that amongst his many musings was a theory of pangenesis involving transmission of acquired traits (Darwin, 1868).

[4] In biology, the term 'gene' generally refers to a sequence of DNA or RNA nucleotides that code for a molecule with a particular function. It will be argued that, with respect to cultural evolution, there is no self-assembly code, and thus no equivalent to the gene.

[5] We are not referring here to clock time but to the *relative* mean duration of biological versus cultural generation processes.

offspring—by way of a self-assembly code (e.g., DNA), and therefore not obliterated at the end of a generation. This is not the case with respect to cultural traits (Gabora, 2011b). (Nor, as we shall see, is it even the case for all biological traits.)

As a final preliminary note, it is important to keep in mind that organisms (including humans) are affected by epigenetic processes that influence the regulation and expression of genetic traits due to interactions with the environment, as well as selection effects operating on groups as well as individuals (Wilson, 1975). For simplicity, this paper does not explore these complications in detail but their relevance to the argument presented here is discussed elsewhere (Voorhees, Read, & Gabora, in press).

## Randomness

It is possible for a selectionist model to be applicable even if the underlying process is not random, but in that case, although not genuinely random, the process must be approximated by a random distribution.[6] Biological variation is not genuinely random (for example, we can trace the source of some mutations to various causal agents; see Caporale 2000) but the assumption of randomness generally holds sufficiently well to serve as a useful approximation.[7]

With respect to culture, variation is not randomly generated, nor can it be described by a random distribution. Selectionist cultural theorists sometimes concede this point (Heyes, 2018), but fail to recognize its implications for the assumed validity of a selectionist framework. Natural selection *acts upon* nonrandomly generated variation, but to the extent that variation is *not* randomly generated, the distribution of variants reflects whatever is biasing the generation away from random in the first place, rather than selection (i.e., differential selection on the distribution of randomly generated heritable variation in a population over generations). Let us break this argument down step by step.

1. Natural selection is a two-step process, consisting of (i) generation of random variants that differ in fitness, followed by (ii) differential survival and reproduction of the fittest variants.
2. The first step provides variation upon which selection can operate, and the adaptiveness of the process resides not in the first step (how variants are *generated*) but the second (how fit variants are *selected*).
3. To the extent that variants do not differ in fitness, their evolution is attributed not to selection but to random genetic drift (Fisher, 1930; Hartl & Clark, 2006).[8]
4. To the extent that the generation of variants cannot be described by a random distribution, their evolution is attributed not to selection but to the nature of this nonrandom generation process.
5. Cultural change cannot be approximated by a random distribution; it is orders of magnitude more non-random than biological evolution. It is strategic and creative, with ideas emerging due to spreading activation and overlap amongst distributed mental representations encoded in associative memory (Gabora, 2011a).
6. Therefore, a selectionist model is inappropriate to the description of cultural change.

In the cultural Darwinism literature there is much discussion of *social learning* (obtaining *existing* information from someone else), and some mention of *individual learning* (obtaining *existing* information through nonsocial means), but little about creativity, reasoning, planning, problem solving, i.e., the the highly non-random higher cognitive processes that *generate* cultural novelty. In a paper titled "grand challenges for the study of cultural evolution" (Brewer et al., 2017), absent from among the eight challenges is the challenge of studying the creative processes that fuel cultural evolution. The closest they come is to ask "How are innovations selectively transmitted" and "Do innovations create feedback loops leading to cumulative culture?" It seems that understanding how innovations come about in the first place is more fundamental than knowing how they are "selectively transmitted" or whether they create feedback loops. Without the creative generation of cultural novelty, there is no cultural evolution. As demonstrated in an agent-based computational model of cultural evolution (Gabora, 1995); when agents never *imitate,* cultural evolution does occur, albeit slowly, as each agent figures things out on its own, but when agents never *create,* there is no cultural evolution at all. Thus, understanding creativity would appear to be the 'grandest' challenge of all for cultural evolution research.

The 'randomness' argument puts a major dent in the theory that cultural evolution is selectionist, but it does not destroy it altogether. It is possible that after variation has been generated by way of nonrandom processes there might still be work for selection to do in winnowing out the very fittest. However, we now turn to the more serious problem, that in cultural evolution there is no self-assembly code.

## Self-assembly Code

In biological evolution there are two kinds of traits: (1) inherited traits (e.g., blood type), transmitted vertically from parent to offspring by way of genes, and (2) acquired traits (e.g., a tattoo), obtained during an organism's lifetime, and transmitted horizontally amongst conspecifics.[9] A selectionist explanation works in biology to the extent that retention of acquired change is negligible compared to retention of selected change; otherwise the first, which can operate instantaneously, overwhelms the second, which takes generations. Transmission of acquired traits is avoided through use of a *self-assembly code* (such as the genetic code), i.e., a set of instructions for how to reproduce. Because a lineage perpetuates itself using a self-assembly code, inherited traits are transmitted while acquired traits are not.[10]

---

[6] Few things other than radioactive decay are truly random.

[7] Actually, in some biological situations, such as assortative mating, the assumption of randomness does not hold, and in such cases natural selection is not an appropriate model.

[8] Drift has been demonstrated in human culture (Bentley, Hahn, & Shennan, 2004), and in a computational model of cultural evolution (Gabora, 1995).

[9] This is a simplification, for there exist traits that are encoded in genetic material, but this genetic material does not constitute a full-fledged self-assembly code (see Bondurianksy & Day, 2009).

Now let us turn to culture. In cultural evolution, there is no self-assembly code, and no vertically transmitted inherited traits; all change is acquired.[11] Therefore, cultural evolution is not due to the mechanism Darwin proposed: differential replication of heritable variation in response to selection. The only response to this argument I know of comes from Mesoudi (2007): "[the] point concerning the lack of self-assembly codes in cultural entities is, again, well-taken when compared to many biological organisms, but may not hold if we take viruses as our biological exemplar, which similarly cannot self-replicate in the absence of a host, or … the evolution of early RNA-based life before DNA-based replication mechanisms evolved." This response evades the problem, for the argument is not that cultural evolution differs from biological evolution, but that the assumptions underlying the formal framework developed to describe evolution by natural selection renders it inapplicable to culture. Indeed, it is also inapplicable to the description of some aspects of biological evolution, but that should be more reason for concern, not less.

Thus, to help determine whether there is a genuine flaw in the argument, and if so pinpoint what that flaw is, we again break the argument down into steps:

1. To the extent that an evolutionary process is amenable to a selectionist model, there are two kinds of traits: vertically transmitted inherited traits, and horizontally transmitted acquired traits.
2. Acquired traits are discarded from a lineage at the end of every generation.
3. This means that evolution (i.e., cumulative, open-ended, adaptive change) in biological lineages cannot be explained in terms of acquired traits.
4. Therefore, it is explained in terms of inherited traits.
5. In biological evolution, inherited traits are not discarded; they are preserved by way of a self-assembly code. The code's low-level information-bearing components must be organized in an orderly manner so they can be parsed into meaningful units; otherwise, the precisely orchestrated process by which the code is expressed to generate offspring is disrupted.
6. The population genetics framework was developed to explain change in a system such as this where the slow process of selection for inherited traits over generations is not drowned out by the fast process of acquired change (which can take place over milliseconds).
7. Biological evolution is therefore explainable in terms of differential selection on the distribution of randomly generated heritable variation in a population over generations, i.e., natural selection.
8. Since acquired change operates markedly faster than inherited change, to the extent that acquired change is *not* wiped out at the end of each generation, a population genetics framework is inappropriate as an explanatory model.
9. In cultural evolution, there is no distinction between vertically transmitted inherited traits and horizontally transmitted acquired traits. Since all traits are horizontally transmitted, we may refer to them as *cultural acquired traits*.
10. Cultural acquired traits are *not* regularly discarded from cultural lineages at the end of generations.
11. This means that evolution (i.e., cumulative, open-ended, adaptive change) in cultural lineages *can* be explained in terms of acquired traits.
12. Moreover, culture is not transmitted by way of inherited traits.
13. Therefore, cultural change, unlike biological change, cannot be explained in terms of change in the frequency of inherited traits; there exists no basis upon which to explain cultural evolution in terms of differential selection of inherited traits on the distribution of randomly generated heritable variation in a population over generations, i.e., using a selectionist framework.
14. Cultural evolution must therefore be explained entirely in terms of changes to acquired traits.

This argument has important implications for how cultural data is modeled. Since biological acquired traits are usually (though not always) discarded, and since a self-assembly code must stay intact to preserve its self-replication capacity, the joining of bifurcations in biological lineages is rare; thus, a phylogenetic tree correctly captures the branching structure. However, since cultural acquired traits are not discarded, and there is no cultural self-assembly code, the joining of bifurcations in cultural

---

[10] An organism may bypass the disappearance of acquired traits through niche construction, *i.e.,* by modifying its environment in such a way as to influence the behavior (and potentially gene regulation) of offspring. Thus, by 'building acquired traits into the environment', one generation may influence the traits exhibited by the next (Lewontin, 1998). However, acquired change is sufficiently negligible relative to inherited change that a selectionist explanation is still of value in explaining biological evolution.

[11] An anonymous reviewer suggested natural language is a cultural self-assembly code. However, (1) natural language is not a set of encoded instructions for the self-replication of natural languages, and (2) culture does not exhibit the signature characteristics of evolution by way of a self-assembly code: lack of transmission of acquired traits, and culture is characterized by horizontal not vertical transmission. Nevertheless, culture may be moving *toward* a cultural Darwinian threshold. In other words, it may exist in the state biological life was in before LUCA (last universal common ancestor) (Woese, 1998).

'lineages' is commonplace, and thus the structure is network-like rather than the tree-like (Gabora, 2006b). This difference has been demonstrated mathematically using split-decomposition graphs (Bandelt & Dress, 1992; Wägele, 2005). Dress and colleagues showed that while biological data generate branching graphs, reanalysis of data from a psychological experiment in which people were asked to estimate the subjective distance between colours gives a very different structure (Dress, Huson, & Moulton, 1996). This difference in the deep structure of biological data and cultural data such as languages, concepts, and artifacts arising from human cognition, is why phylogenetic tree models of culture are problematic.

## Self-Other Reorganization (SOR): An Alternative Approach to Cultural Evolution

The above analysis precludes a *selectionist* but not an *evolutionary* framework for culture. Indeed, research since the Modern Synthesis has shown that even life itself is only partially explained through recourse to a selectionist framework; for example, though biological traits are generally obtained through vertical inheritance, horizontal gene transfer (HGT) involves horizontal transmission (Ochman et al., 2000). Evolution can occur in the absence of selection, and the importance of non-selectionist processes in evolution is increasingly recognized (Kauffman, 1993; Killeen, 2017; Woese, 2002). Research on the origin of life suggests that early life consisted of autocatalytic protocells that evolved through a non-selectionist means, and natural selection emerged later from this more haphazard, ancestral evolutionary process (Baum, 2018; Cornish-Bowden & Cárdenas, 2017; Gabora, 2006; Goldenfeld, Biancalani, & Jafarpour, 2017; Hordijk, Steel, & Dittrich, 2018; Steel, 2000; Vetsigian, Woese, & Goldenfeld, 2006). This non-selectionist process requires (1) a *self-organizing network* of components that generate new components through their interactions, (2) the network should be able to reconstitute another like itself through haphazard (not code-driven) duplication of components, and (3) interaction amongst networks. This process can be referred to as *Self-Other Reorganization* (SOR) because it involves an interplay between self-organized *internal* restructuring, and communal exchange *amongst* autocatalytic structures. Change occurs not through selection but through a process that has a completely different mathematical description: context-driven actualization of potential (Gabora & Aerts, 2005). The entity changes through interactions with its world, which in turn alters its potential for future configurations. Like selectionist evolution, SOR has mechanisms for preserving continuity and for introducing novelty, but unlike selectionist evolution, it is a low-fidelity Lamarckian process. The distinction between these two processes is summarized in Table One.

Vetsigian et al. (2006) showed that to cross the Darwinian threshold from non-selectionist to selectionist evolution requires the emergence of a self-assembly code. There is no evidence that culture has crossed this threshold, and it does not possess the *sine qua non* of having crossed it: vertical transmission and a lack of transmission of acquired traits. It has been proposed that, as did early life, culture evolves through SOR (sometimes referred to as 'communal exchange') (Gabora, 1999, 2004, 2019). Here, the self-organizing networks are not protocells exchanging catalytic molecules, but minds exchanging ideas. Tools improve and fashions change not through selection but through context-driven actualization of potential. As parents and others share knowledge with children, an integrated network of understandings takes shape in their minds, and they become creative contributors to cultural evolution. It has been noted that a tension exists between cultural evolution theory and the literature on human nature (Lewens, 2017). Because SOR is not incompatible with the transmission of acquired traits, and because it recognizes the integrated, 'self-mending' nature of an individual mind, it provides a natural means of reconciling cultural evolution and human nature. That said, SOR is but one of a class of network-based approaches (e.g., Bentley & Shennan, 2003), and other non-selectionist models of cultural evolution, such as those based on the Price equation (e.g., El Mouden, André, Morin, & Nettle, 2014).

Table One: Summary of the distinction between evolution through selection and evolution through Self-Other Reorganization.

|  | **Selection** | **Self-other Re-organization (SOR)** |
| --- | --- | --- |
| Unit of self-replication | Organism | Self-organizing autocatalytic network |
| Mechanism for preserving continuity | Reproduction (vertical transmission), proofreading enzymes, etc. | Communal exchange (horizontal transmission) |
| Generation of novelty | Mutation, recombination | Creativity, transmission error |
| Self-assembly code | DNA or RNA | None |
| High fidelity | Yes | No |
| Transmission of acquired traits | No | Yes |
| Type | Selectionist | Lamarckian (by some standards) |
| Evolution processes it can explain | Biological | Early life; horizontal gene transfer, culture |

## Conclusions

Darwin faced the problem of explaining how lineages evolve despite that acquired changes are lost from a lineage when the individuals that acquired them dies. Darwin's solution was to come up with a population-level (macro) explanation. His theory of natural selection holds that although *acquired traits* are discarded, *inherited traits* are retained, so evolution can be explained in terms of preferential selection for those inherited traits that confer fitness benefits on their bearers. Cultural evolution research does not face the problem that motivated Darwin's solution—that of explaining how evolution takes place despite the discarding of acquired traits—because cultural acquired traits are *not* discarded. Thus, while the rationale for framing culture as an *evolutionary* process is correct, it does not follow that culture is a selectionist process, or that population genetics provides viable starting points for modeling cultural change. Cultural evolution research has been carried out largely independent of research on the mental structures that actually evolve culture. This has led to the mis-application of biological constructs such as generations, inheritance, and vertical / horizontal transmission. This in turn has hindered appreciation of the dependence of vertical inheritance on a self-assembly code, and recognition of the implications of its absence in cultural evolution. The field is in need of cognitive scientists to uncover the cognitive processes by which culture actually takes shape.

Psychologists use the term mental set to refer to the persistent use of problem-solving strategies that worked in the past even when these strategies are not appropriate to the problem at hand. It appears that the persistent application of a selectionist framework to cultural evolution, despite that the conditions that make that framework applicable in biology are not present with respect to culture, may be an instance of mental set. This paper has laid out step-by-step arguments as to why a selectionist approach to culture is inappropriate, and pointed to an alternative approach.

## Acknowledgments

This work was supported by a grant (62R06523) from the Natural Sciences and Engineering Research Council of Canada.